\documentclass[11pt]{article}
\usepackage{amsfonts}
\usepackage{bm}
\usepackage{amsmath}
\usepackage{epsfig}
\usepackage[latin1]{inputenc}
\usepackage{amssymb,lscape}
\usepackage[matrix,arrow,curve]{xy}

\newcommand{\bmat}{\left(\begin{array}}
\newcommand{\emat}{\end{array}\right)}

\def\gtrsim{\mathrel{\raise.3ex\hbox{$>$\kern-.75em\lower1ex\hbox{$\sim$}}}}

\def\-{\hphantom{-}}

\def\s2{\frac{1}{\sqrt2}}

\def\mg{m_{3/2}}
\def\mg2{m^2_{3/2}}

\def\Dsl{\,\raise.15ex\hbox{/}\mkern-13.5mu D} 

\def\be{\begin{equation}}
\def\ee{\end{equation}}
\def\bea{\begin{eqnarray}}
\def\eea{\end{eqnarray}}

\begin{document}

\pagestyle{plain}

\makeatletter
\@addtoreset{equation}{section}
\makeatother
\renewcommand{\theequation}{\thesection.\arabic{equation}}
\pagestyle{empty} 
\begin{flushright}
DFPD-2014/TH/08
\end{flushright}
\begin{center}
\LARGE{ \bf Gaugings from $E_{7(7)}$ Extended Geometries \\[10mm]}
\large{ Walter H. Baron\\[6mm]}
\small{ {\em Dipartimento di Fisica e Astronomia ``Galileo Galilei'', \\
Università di Padova, Via Marzolo 8, 35131 Padova, Italy \\[6mm] 
INFN, Sezione di Padova,\\ Via Marzolo 8, 35131 Padova, Italy}.\\[6mm]
wbaron@pd.infn.it}\\[2cm]

\small{\bf Abstract} \\[0.5cm]\end{center}
We discuss the construction of gaugings in recent models of $E_7$ Extended Geometries, focusing on the two inequivalent $SL(8)$ truncations of the theory. In these sectors the conditions for the generation of gaugings in the ${\bf 36, 36', 420}$ and ${\bf 420'}$ representations of $E_{7(7)}$ can be compactly expressed in terms of objects which are in the fundamental representation of $SL(8)$, making the search of solutions simpler. As an application we discuss the generation of $SO(8)$ gaugings. In particular we show how the internal generalized vielbein for the seven sphere recently found by Nicolai {\it et al.} can be derived in a completely independent setting and we also prove that neither of these sectors is able to generate the new $SO(8)$ dyonic gaugings, at least if the so called section conditions are implemented.  
\\[4cm]

\newpage
\setcounter{page}{1}
\pagestyle{plain}
\renewcommand{\thefootnote}{\arabic{footnote}}
\setcounter{footnote}{0}

\tableofcontents

\section{Introduction}

There has been a strong interest in Gauged supergravities over the last years and the success in describing these in a unified way is certainly thanks to the development of the embedding tensor formalism \cite{Nicolai:2001sv,deWit:2002vt,de Wit:2007mt}. Several attempts have been done to explore if these theories have or not an uplift to higher dimensional Supergravity or String/M Theory. One notable example, in the 4D case, is the $SO(8)$ gauged supergravity which can be obtained by compactification of 11D supergravity on $AdS_4\times S^7$ \cite{de Wit:1986iy,Nicolai:2011cy}. Recently it was shown that new inequivalent maximal supergravities with $SO(8)$\footnote{It also applies for the non-compact gauge groups $CSO(p,q,r)$, with $p+q+r=8$.} symmetry group can be generated \cite{DallAgata:2011aa,Dall'Agata:2012bb}. These theories can be seen as a deformation of the original one by a parameter $c$ which measures the way in which the electric and magnetic potentials couple to the generators of the gauge algebra, so these models are usually denoted as the $SO(8)_c$ gauged supergravities. 

Attempts to embed these theories in 11D supergravity have failed (see $e.g.$ \cite{deWit:2013ija}), and so a natural exercise is to explore whether these theories can be reproduced within the U duality covariant extensions of 11D supergravity proposed in the literature, as for instance the Extended Geometry (EG) formalism of  \cite{Aldazabal:2013mya,Aldazabal:2013via} and the Exceptional Field Theory (EFT) of \cite{Hohm:2013pua,Hohm:2013vpa,Hohm:2013uia}. These proposals are based on the formalism developed in \cite{Coimbra:2011ky,Coimbra:2012af,Berman:2012vc}. The EFT is properly defined once the so called {\it section conditions} are imposed (see section \ref{Implications}) while the EG allows for possible relaxation of them\footnote{See the discussion in \cite{Aldazabal:2013mya} for the 7 dimensional situation concerning us and also \cite{Berman:2012uy,Musaev:2013rq} for the particular situations where the internal space is 4, 5 or 6 dimensional.}. In EG the analysis is focused in geometric aspects of the extended space, and the discussion is centered in the generalized diffeomorphisms and their gauge consistency conditions. In contrast, in the EFT approach a precise theory with an action in 56 + 4 dimensions was developed. The building of gaugings by mean of generalized vielbeins and generalized Scherk Schwarz reductions is based on the internal generalized diffeomorphisms which is the link among all these approaches, so the discussion in this work remains general.

The connection of these works with 11D supergravity can be done much more simpler with the new $E_7$ covariant reformulation developed in \cite{deWit:2013ija,Hillmann:2009ci,Godazgar:2013dma,Godazgar:2013pfa,Godazgar:2013oba,Godazgar:2014sla}, where the internal degrees of freedom, comprising the metric, three form and six dual forms are accommodated in a generalized 56-bein taking values in $E_7/SU(8)$. This generalized frame arises from the study of the supersymmetry transformation of both the electric and the magnetic vector fields and the internal generalized vielbein can be read out by using the non-linear metric and flux ansatz. Henceforth we will deal only with the internal space, thus unless otherwise specified, when we write ``generalized vielbein'' we will refer to the internal piece.

One of the great potentialities in these frameworks rests in that they provide a direct mechanism for the search of gaugings. Each embedding tensor defines a system of differential equations which, {\it a priori}, can be integrated out if the solution is interpretable as a generalized Twisted Torus. Implementing this in the practice could be excessively involved, but the exploration can be made simpler when particular sectors of the original structure are considered. In this work we address this situation for an internal bein constrained to one of the maximal subgroups of $E_7$. We want to stress here that a related analysis was performed in DFT in \cite{Dibitetto:2012rk}. There the authors considered the explicit construction of gaugings in term of generalized frames and were able to perform an exhaustive exploration by restricting to the situation where the external space time was $d\ge7$. Interestingly enough they built $SO(4)_c$ gaugings in the $d=7$ case with a truly $3+3$ doubled solution.

Of course, once a generalized bein reproducing some particular gauging is obtained, one should confirm the consistency of the solution with the corresponding field equations. 

This paper is organized as follows. In section \ref{secET} we pose the Extended Geometry setup and describe its connection with the embedding tensor formalism. In sections \ref{secSL8} and \ref{secSL8'} we separately perform the truncation\footnote{In this work we focus exclusively in the internal space, hence by ``truncation'' we understand an exploration of gaugings for some subset of the full configuration space of the generalized internal beins and it does not make any reference to the scalar sector, so it must be not confused with a truncation of a Theory.} 
to the two SL(8) subgroups of $E_7$ and we compute the components of the embedding tensor in the $sl(8)$ branching. We then probe the usefulness of these expressions by making an exhaustive exploration of $SO(8)_c$ gaugings withing these sectors, in particular we show how to recover the $SO(8)$ electric ones and confirm that there are no dyonic gaugings, at least if the section condition is not relaxed. In section \ref{secConcl} we review the results and comment on possible future extensions of this work. Some technical details about $E_{7(7)}$ and gamma matrices are left to the Appendix.

\section{Embedding tensors and internal fluxes in $E_7$ Extended Geometries}\label{secET}

In the U duality covariant approach of \cite{Aldazabal:2013mya} the generalized vielbein $E_{\mathbb A}{}^{\mathbb M}$ of the megatorus is valued in ${\mathbb R}^+\times E_{7(7)}/SU(8)$. It is parametrized by a conformal factor $\Delta$ and an $E_7$ frame $U_{\mathbb A}{}^{\mathbb M}$,
\bea
E_{\mathbb A}{}^{\mathbb M}= e^{-\Delta}~ U_{\mathbb A}{}^{\mathbb M}.
\eea 

$E_{\mathbb A}{}^{\mathbb M}$ can be seen as a vector, transforming under the generalized diffeomorphisms \cite{Coimbra:2011ky,Berman:2012vc}
\bea
\delta_\xi V^M=\xi^P \partial_P V^M-12 P_{(adj)}{}^M{}_N{}^P{}_Q \partial_P\xi^Q V^N + \frac\omega2 \partial_P\xi^P V^M~, \label{GenDiff}
\eea
encoding both diffeomorphisms and gauge transformations. The first two terms in (\ref{GenDiff}) guarantee the preservation of the $E_7$ structure and the last term comes from the ${\mathbb R}^+$ transformations, the weight $\omega$ is 1 for $E$ but vanishes for the $E_7/SU(8)$ bein $U$ and $P_{(adj)}$ denotes the projector on the adjoint representation (see (\ref{padj}) ). Indices are raised and lowered by using the NorthEast-SouthWest convention with the weighted simplectic matrices $\omega_{{\mathbb M}{\mathbb N}}=-ie^{2\Delta}\sigma_2\otimes {\mathbb I}_{28}$ and $\omega^{{\mathbb M}{\mathbb N}}=e^{-2\Delta}\omega_{{\mathbb M}{\mathbb N}}$ for curved indices and similarly for the matrix with flat indices $\omega_{{\mathbb A}{\mathbb B}}$ obtained by setting $\Delta=0$ above. In particular these conventions imply $E^{\mathbb A}{}_{\mathbb M}=-\left(E^{-1}\right)_{\mathbb M}{}^{\mathbb A}$ . 

These transformations define the generalized fluxes, $F_{{\mathbb A} {\mathbb B}}{}^{\mathbb C}$, via 
\bea
\delta_{E_{\mathbb A}}E_{\mathbb B}= F_{{\mathbb A} {\mathbb B}}{}^{\mathbb C} E_{\mathbb C}~.\label{Fdef}
\eea
They determine the gaugings of the Effective Theory, when the megatorus is seen as the internal space of a higher dimensional theory. Equation (\ref{GenDiff}) leads to
\bea
F_{{\mathbb A} {\mathbb B}}{}^{{\mathbb C}}=X_{{\mathbb A} {\mathbb B}}{}^{{\mathbb C}}+D_{{\mathbb A} {\mathbb B}}{}^{{\mathbb C}},
\eea
\bea
X_{{\mathbb A} {\mathbb B}}{}^{{\mathbb C}}=\Theta_{\mathbb A}{}^{\alpha}\left[t_\alpha\right]_{\mathbb B}{}^{\mathbb C}~,~~~~~~\
\Theta_{\mathbb A}{}^{\alpha}=7P_{(912)}{}_{\mathbb A}{}^{\alpha}{},^{B}{}_{\beta}\tilde\Omega_{\mathbb B}^{\beta},\label{Xtensor}
\eea
\bea
D_{{\mathbb A} {\mathbb B}}{}^{{\mathbb C}}=-\vartheta_{\mathbb A}\delta_{\mathbb B}^{\mathbb C}+ 8~P_{(adj)}{}^{\mathbb C}{}_{\mathbb B}{}^{\mathbb D}{}_{\mathbb A}\vartheta_{\mathbb D}~,~~~\vartheta_{\mathbb A}=-\frac12 \left(\tilde\Omega_{{\mathbb B}{\mathbb A}}{}^{\mathbb B}-3\partial_{\mathbb A}\Delta\right)~,
\eea
wherein $X_{{\mathbb A} {\mathbb B}}{}^{{\mathbb C}}$ and $D_{{\mathbb A} {\mathbb B}}{}^{{\mathbb C}}$ are the projections in the ${\bf 912}$ and ${\bf 56}$ representations, and $\tilde \Omega_{{\mathbb A} {\mathbb B}}{}^{{\mathbb C}}$ is the flat index Weitzenböck connection of the $E_7$ piece, 
\bea
\tilde\Omega_{{\mathbb A}{\mathbb B}}{}^{{\mathbb C}}=e^{-\Delta}~ U_{\mathbb A}{}^{\mathbb M}~ U_{\mathbb B}{}^{\mathbb N}~ \partial_{\mathbb M} U^{\mathbb C}{}_{\mathbb N}~,\label{Weitzenbock}
\eea 
and is an element of the algebra of $E_{7}$, $\tilde\Omega_{{\mathbb A}{\mathbb B}}{}^{{\mathbb C}}=\tilde\Omega_{\mathbb A}{}^\alpha\left[t_\alpha\right]_{\mathbb B}{}^{\mathbb C}$. $P_{(912)}$ in (\ref{Xtensor}) is the projector on the {\bf 912} representation of $E_7$ and is defined in (\ref{p912}).

It is interesting to note that from the definition above, the structure constants $X_{\mathbb A\mathbb B\mathbb C}$ automatically satisfy the 4D maximal supergravity relations of the embedding tensor \cite{de Wit:2007mt}
\bea
P_{(adj)}{}^{\mathbb C}{}_{\mathbb B}{}^{\mathbb D}{}_{\mathbb E}~X_{\mathbb A\mathbb D}{}^{\mathbb E}= X_{\mathbb A\mathbb B}{}^{\mathbb C}~ ,
~~~X_{\mathbb A[\mathbb B\mathbb C]}=X_{\mathbb A\mathbb B}{}^{\mathbb B}=X_{(\mathbb A\mathbb B\mathbb C)}=X_{\mathbb B\mathbb A}{}^{\mathbb B}= 0.
\eea
The consistency of the extended geometry setup is guaranteed if all fields and gauge parameters (denoted ${\mathcal A},{\mathcal B}$ below) are in a particular seven dimensional section of the original 56-megatorus. These restrictions are imposed by $E_7$ covariant constraints known as {\it section conditions},
\bea
\Omega^{{\mathbb M} {\mathbb N}}\partial_{{\mathbb M}}{\cal A}~\partial_{\mathbb N}{\cal B}=0,~~
[t_\alpha]^{{\mathbb M}{\mathbb N}}\partial_{{\mathbb M}}{\cal A}~\partial_{\mathbb N}{\cal B}=0,~~~[t_\alpha]^{{\mathbb M}{\mathbb N}}\partial_{{\mathbb M}}\partial_{\mathbb N}{\cal A}=0~,\label{SecCond}
\eea
and are the analogous of the {\it strong constraint} in Double Field Theory (DFT).

Notice that the choice of physical section $X^{m8}=y^m,m=1,\dots,7$ reduces the Extended Geometry formulation reviewed above to the $E_{7(7)}\times {\mathbb R}^+$ Generalized Geometry formulation of \cite{Coimbra:2011ky,Coimbra:2012af}.

We stress here that the {\it section conditions} are sufficient but not necessary conditions in the EG approach. Indeed, consistency is guarantied as long as closure constraint are imposed leading to the quadratic constraint after compactification \cite{Aldazabal:2013mya}. Similar relaxations were found in the Extended Geometries with $E_4$, $E_5$ and $E_6$ exceptional groups \cite{Berman:2012uy,Musaev:2013rq}.

So far there are no known concrete examples where this constraint is relaxed and we will not address this interesting point here and leave it for future investigations. Progress in this direction could be crucial in the study of non-geometric solutions, in particular they could shed some light on the so far open question of whether there is or not a higher dimensional uplift of the new $SO(8)$ gauged supergravities found in \cite{DallAgata:2011aa,Dall'Agata:2012bb}. 

In the following we will focus on the cases without fluxes in the $\bf 56$, $i.e.$ we will set $\vartheta_{\mathbb A}=0$.

\section{The $SL(8)$ case}\label{secSL8}

Let us begin by considering the situation where the generalized internal vielbein is in the diagonal $SL(8,{\mathbb R})$. The ${\bf 56}$ representation of this group is generated by exponentiation of 
\bea
\Lambda_{IJ}{}^{KL}=diag(\Lambda^{ij}{}_{kl},\Lambda_{ij}{}^{kl}),
\eea
\bea
\Lambda^{ij}{}_{kl}=\delta^{[i}{}_{[k}\lambda^{j]}{}_{l]}~,~~\Lambda_{ij}{}^{kl}=\delta_{[i}{}^{[k}\lambda_{j]}{}^{l]}.\label{SL(8)}
\eea
The matrix $\lambda^i{}_j$ is in the fundamental representation of $sl(8)$ and $\lambda_i{}^j=-\lambda^i{}_j$.
Hence 
\bea
U^{ij}{}_{kl}=\left[e^{\Lambda}\right]^{ij}{}_{kl}=2 \left[e^{\frac12\lambda}\right]^{[i}{}_{k}\left[e^{\frac12\lambda}\right]^{j]}{}_{l}=2 U^{[i}{}_{k} U^{j]}{}_{l}~,~~~U_{ij}{}^{kl}=2 U_{[i}{}^{k} U_{j]}{}^{l}~,\label{UU}
\eea
where $U^i{}_j$ is in the ${\bf 8}$ representation and $U_i{}^j$ is its inverse transposed. The unimodular character of U is expressed by the following constraint
\bea
U^A{}_P~\partial_{\mathbb M}U_A{}^{P}=0,\label{unimodular}
\eea
which can be proved by using that the Weitzenböck connection is in the algebra of $SL(8)$. There are also two constraints coming from the requirement of the vanishing of ${\bf 56}$ gaugings $\vartheta_{\mathbb A}$,
\bea
U_{[A}{}^{[M}\partial_{MN}\left( U_{B]}{}^{N]}e^{-\frac{3\Delta}2}\right)=0~,~~~U^{[A}{}_{[M}\partial^{MN}\left( U^{B]}{}_{N]}e^{-\frac{3\Delta}2}\right)=0~.\label{no56rep}
\eea
Plugging (\ref{UU}) into (\ref{Xtensor}) and using (\ref{Weitzenbock}) and the formulae of Appendix A as well as the constraints (\ref{unimodular}) and (\ref{no56rep}) one obtains after tedious but straightforward computations the explicit expression for the embedding tensor. However in the analysis we perform in this work it is convenient to consider only its projections on the $sl(8)$ branching, $i.e.$ ${\bf 36,\,36',\,420,}$ and ${\bf 420'}$, computed with $X_{{\mathbb  A}{\mathbb B}{\mathbb C}}=\omega_{{\mathbb C} {\mathbb D}}X_{{\mathbb  A}{\mathbb B}}{}^{{\mathbb D}}$,\footnote{We use the standard splitting of indices $X_{{\mathbb  A}{\mathbb B}{\mathbb C}}=\left(X_{{\mathbb  A}{\mathbb B}}{}^{CD},X_{{\mathbb  A}{\mathbb B}{CD}}\right)$ with $C,D=1,\dots,8$ and $C<D$, and similarly for ${\mathbb A},~{\mathbb B}$.}
\bea
\theta_{AB}&:=&\frac4{21} X_{AC\,BD}{}^{CD}=-16~e^{-\Delta}~ U_{(A}{}^{[M}~\partial_{MN} U_{B)}{}^{N]}~,\cr
\xi^{AB}&:=&\frac4{21} X^{AC\,BD}{}_{CD}=16~e^{-\Delta}~ U^{(A}{}_{[M}~\partial^{MN} U^{B)}{}_{N]}~, \cr
\zeta_{A}{}^{BCD}&:=& 2~ X_{AE}{}^{BCDE}= 2\cdot4!~e^{-\Delta}\left( ~U^{[B}{}_M U^C{}_N U^{D]}{}_P~\partial^{MN} U_A{}^P -\frac13~\delta_{A}^{[B}~ U^{C}{}_{[M}\partial^{MN} U^{D]}{}_{N]} \right)~,\cr
\eta^{A}{}_{BCD}&:=& 2~ X^{AE}{}_{BCDE}= -2\cdot4!~e^{-\Delta}\left( ~U_{[B}{}^M U_C{}^N U_{D]}{}^P~\partial_{MN} U^A{}_P -\frac13~\delta^{A}_{[B}~ U_{C}{}^{[M}\partial_{MN} U_{D]}{}^{N]} \right)~.\cr\label{SL8gaugings}
&&
\eea
Then we can compactly state the following requirements for the building of gaugings with generalized vielbeins $SL(8,{\mathbb R})$ valued:
\begin{itemize}
\item ${\bf 36}$ (${\bf 36'}$) gaugings require $\partial_{MN}\neq0$ ($\partial^{MN}\neq0$),
\item ${\bf 420}$ (${\bf 420'}$) gaugings require $\partial^{[|MN|}\left(e^{-\frac12 \Delta} U_A{}^{P]}\right)~\neq0$ ($\partial_{[|MN|}\left(e^{-\frac12 \Delta}U^A{}_{P]}\right)~\neq0$),
\end{itemize}
where the bars denote $\partial_{|MN|}=\partial_{MN}$ ($-\partial_{NM}$) if $M<N$ ($M>N$) and we have made used of (\ref{no56rep}).

\subsection{SO(8) gaugings}

As a practical application of (\ref{SL8gaugings}) we consider here the $SO(8)$ gauged supergravity. Dyonic gaugings have vanishing {\bf 420} and {\bf 420'} projections and $\xi=c~\theta^{-1}$, with $\theta_{AB}=\delta_{AB}$ up to rescaling and $SL(8)$ rotations. The purely electric theory ($c=0$) has an uplift to 11D supergravity which can be embedded in our formalism by choosing a section with dependence on $X^{m8}=y^m$, the coordinates over $S^7$. The full generalized vielbein can be read out from \cite{Hillmann:2009ci} or equivalently from the ground up approach from 11 dimensional supergravity\footnote{We are very grateful to H. Nicolai and M. Godazgar for private communications clarifying different aspects of this reformulation.} \cite{Godazgar:2013dma}. But the internal seven dimensional piece leading to the fluxes can be disentangled only after the use of the {\it non-linear ansatz} for the metric \cite{de Wit:1986iy}, the three \cite{deWit:2013ija} and the dual six forms \cite{Godazgar:2013dma}. Later on it was shown that this solution can be embedded in the $E_{7(7)}\times{\mathbb R}^+$ Generalized Geometry formulation in \cite{Lee:2014mla}. In this case, the vanishing of the four form fluxes guarantees that the generalized internal frame is indeed in the $SL(8)$ subgroup considered in this section.

From the Extended Geometry point of view we see that $\partial^{MN}=0$ guarantees the vanishing of the projections of the embedding tensor on ${\bf 36'}$ and ${\bf 420}$.  Then the conditions to be satisfied for the generalized vielbein in order to reproduce the $SO(8)$ gaugings are
\bea
\partial_{[m}\left(e^{-\frac12\Delta}U^A{}_{n]}\right)=0~,~~~~~e^{-\Delta}~ U_{(A}{}^{[m}~\partial_{m} U_{B)}{}^{8]}=\lambda~ \delta_{AB}~,~\label{SO(8)}
\eea
with $\lambda$ being a constant. The first equation above guarantees the vanishing of the ${\bf 420'}$ gaugings and the latter is simply the statement that we are in the particular frame of $SO(8)$ gauged theory where $\theta_{AB}\propto\delta_{AB}$. In addition one has to satisfy (\ref{no56rep}).

Solving this system of equations then leads to an independent derivation of the previously obtained generalized internal vielbein achieved directly from 11D supergravity, illustrating the efficiency of this formalism. In the ground up approach of Nicolai {\it et al.} the generalized internal vielbein and {\it non-linear flux ansatz} were deduced after involved analysis connecting both the bosonic and the fermionic sectors. In fact a central role was played by the supersymmetry transformations of both the electric and the magnetic potentials. Instead here we show how to obtain this by following the opposite direction, $i.e.$ we start with an embedding tensor and deduce the system of equations which determines the generalized vielbein. 

So, our aim here is not to determine explicitly the solution, which was shown to exist,\footnote{As we commented above the Extended Geometry formalism reduces to the Generalized Geometry approach of \cite{Coimbra:2011ky,Coimbra:2012af} after choosing the section along $X^{m8}$ and it was verified in \cite{Lee:2014mla} that the solution of Nicolai {\it et al.} in fact satisfies the algebra (\ref{Fdef}) with the $SO(8)$ embedding tensor.} but simply to obtain the integrable system of equations leading to it. 

The first important observation is that (\ref{SO(8)}) are not independent equations, in fact multiplying the second equation above by $U^B{}_n$ and using
\bea
U_{(A}{}^{[m}~\partial_{m} U_{B)}{}^{8]} &=& U_{B}{}^{[m}~\partial_{m} U_{A}{}^{8]}-U_{[B}{}^{[m}~\partial_{m} U_{A]}{}^{8]}\cr
                                         &=& U_{B}{}^{[m}~\partial_{m} U_{A}{}^{8]}-\frac32 U_{B}{}^{[m}~U_{A}{}^{8]}\partial_{m} \Delta,
\eea
we get
\bea
U^A{}_{m}= \frac{\lambda^{-1}}{2}~e^{\frac12\Delta}~\partial_m \left(e^{-\frac32\Delta}~ U_A{}^8\right),\label{Um}
\eea
which trivially satisfies the first equation in (\ref{SO(8)}). Similarly we can multiply (\ref{SO(8)}) by $U^B{}_8$ to obtain
\bea
U^A{}_{8}= -\frac{\lambda^{-1}}{2}~e^{\frac12\Delta}~\partial_m \left(e^{-\frac32\Delta}~ U_A{}^m\right).\label{U8}
\eea

To solve these equations notice that (\ref{no56rep}) reduces to
\bea
\partial_{m}\left(U_{[A}{}^m U_{B]}{}^8e^{-3\Delta}\right)=0,\label{Kill}
\eea
and there is a very natural ansatz to solve this equation. In fact, from Kaluza Klein reduction we expect the U twist to be a function of geometric objects, as for instance the Killing vectors, and it is well known that the latter are solutions of a similar equation. In fact, given a particular background with $N$ Killing vectors $K_\alpha^m$, $\alpha=1,2,\dots,N$, these satisfy
\bea
\partial_m\left( \sqrt g~K^m_\alpha\right)=0.
\eea 
In the $S^7$ space there are $28$ Killing vectors, which can be parametrized with the antisymmetric pair of indices $AB$.\footnote{If we embed the seven sphere with radius $R$ in the 8 dimensional flat Euclidean space, parametrized by $Y^A,~A=1,\dots,8$, the Killing vectors are simply the angular momentum operators $K_{AB}=\frac1R\left(Y^A\partial_B-Y^B\partial_A\right)$. To make contact with the notation of Nicolai, where the Killing vectors have $SU(8)$ indices $IJ$, one needs and extra gamma matrix, similar to the one used in the next section but with opposite self duality properties.} Hence, one has the ansatz
\bea
U_{[A}{}^m U_{B]}{}^{8}e^{-3\Delta}=\sigma\sqrt{g} K_{AB}^m,\label{KillAnsatz}
\eea
with constant $\sigma$. Multiplying (\ref{KillAnsatz}) with $U^A{}_{n}$ and taking the trace we get
\bea
U_A{}^{8}e^{-\frac32\Delta}=\frac{\kappa}{\lambda} K^m_{AB}~\partial_m \left(U_B{}^{8}e^{-\frac32\Delta}\right)~,\label{UA8}
\eea
wherein $\kappa=\sigma\sqrt{g}~e^{2\Delta}$. Similarly, contracting (\ref{KillAnsatz}) with $U^A{}_8$ leads to 
\bea
U_A{}^{m}e^{-\frac32\Delta}=-\frac{\kappa}{\lambda} K^m_{AB}~\partial_n \left(U_B{}^{n}e^{-\frac32\Delta}\right)~.\label{UAm}
\eea
Setting $\kappa$ to be a constant, equation (\ref{UA8}) (equation (\ref{UAm})) are 8 (56) linear partial differential equations for 8 (56) unknowns $U_A{}^{8}e^{-\frac32\Delta}$ ($U_A{}^{m}e^{-\frac32\Delta}$). Once we have these solutions, $U^A{}_m$ and $U^A{}_8$ can be deduced from (\ref{Um}) and (\ref{U8}).
It is worth mentioning that the explicit solution of Nicolai {\it et al.} indeed satisfies the ansatz (\ref{KillAnsatz}) with constant $\kappa$.

\subsection{Implications of the section conditions}\label{Implications}
Here we make some comments on the consequences of the {\it section conditions} (\ref{SecCond}) over the generalized bein and the construction of gaugings. 

Let us suppose the physical world is parametrized by certain subset of the coordinates $(X^{MN},X_{MN})$, the first section condition in (\ref{SecCond}) requires that half of the coordinates must be absent, but in principle we can choose some $X^{MN}$ and some dual $X_{MN}$. The only restriction is that if fields have a dependence on $X^{ij}$, for certain pair $[ij]$, then the theory must be independent of the dual coordinate $X_{ij}$. 

In order to study the other section conditions we consider separately the cases where $t_\alpha$ is in $sl(8)$ or in its $e_7$ complement. We will refer to these as the $SL(8)$ and $E_7/SL(8)$ section conditions.

Using the expressions for the generators in the Appendix, the latter are shown to be equivalent to 
\bea
\partial^{[|MN|}{\mathcal A}~ \partial^{|PQ|]}{\mathcal B}=0~,~~~
\partial_{[|MN|}{\mathcal A}~\partial_{|PQ|]}{\mathcal B}=0~,\label{E7/SL8constr}
\eea
where the bars denote canonical order as introduced below equation (\ref{SL8gaugings}). If we allow the gauge parameters to depend in an arbitrary way on the section coordinates, we see that (\ref{E7/SL8constr}) again reduces the possible coordinate dependence to a half. Only 7 $X^{MN}$ and 7 duals $X_{PQ}$ survive. The choice must be done in such a way that one of the indices is repeated for the standard and dual coordinates, $e.g.$ possible choices for $X^{MN}$ are $\{X^{m8}\},m=1,\dots,7$ or $\{X^{n7},X^{78}\},n=1,\dots,6$ and so on.

The $SL(8)$ section conditions have severe implications, actually if we look for 7 dimensional sections only 7 $X^{MN}$ or 7 dual coordinates $X_{PQ}$ are allowed. Mixed sections containing both kind of directions are only realizable on lower dimensional truncations. 

To see this, notice that the SL(8) constraint together with $\Omega^{\mathbb M \mathbb N}\partial_{\mathbb M}{\mathcal A}\partial_{\mathbb N} {\mathcal B}=0$
requires
\bea
\partial^{|MP|}{\mathcal A} ~ \partial_{|NP|} {\mathcal B}=0.\label{SL8constr}
\eea
Of course, any 7 dimensional section along $X^{|MQ|}$ or along $X_{|MQ|}$ (with $Q$ fixed and $M=1,\dots,8\neq Q$) is a solution to (\ref{SL8constr}). Regarding mixed sections, these are realizable only when we have dependence on at most six coordinates. These mixed solutions could be relevant in the study of deformations of gauged supergravities coming from ten dimensional supergravities or DFT. 

These requirements have strong consequences in the construction of solutions leading to 4D $SO(8)_c$ gauged maximal supergravities, in fact the expected scenario is that they are continuously connected with the electric solutions and so they should be 7 dimensional. Hence, from (\ref{SL8gaugings}) we see that internal beins taking values in $SL(8,{\mathbb R})$ are only able to build purely electric or purely magnetic gaugings of the $36$ and $36'$. 

This implies that the only hope of getting dyonic gaugings in this sector could be realized from a weakening of the {\it section conditions}. It is worth mentioning here that in the context of DFT, there are well known situations (see \cite{Aldazabal:2011nj,Geissbuhler:2011mx,Grana:2012rr}) where the strong condition was successfully relaxed. The development of examples of this kind in the U-duality extension of DFT is an interesting but nontrivial issue and is beyond the scope of this work. In spite of that we point out that neither the truncation of this section (\ref{SL8gaugings}) nor the upcoming (\ref{SL'8gaugings}) was performed by explicit use of the {\it section conditions}. 
\section{The $SL'(8)$ case}\label{secSL8'}
Another relevant sector of the theory is when the bein is contained in a different maximal subgroup of $E_{7(7)}$. As reviewed in the Appendix, it contain three maximal subgroups, one $SU(8)$ and two $SL(8)$.\footnote{Actually, as they are defined by their action over second rank tensors, these groups correspond to $SU(8)/{\mathbb Z}_2$ and $SL(8)/{\mathbb Z}_2$ respectively.} In order to discriminate among both special linear subgroups we will call them as $SL(8)$ and $SL'(8)$, where the unprimed is the one considered in the previous section.

Again we want to perform a truncation where the fluxes are expressed in a more simple way. The aim is to provide simple sectors of the theory where the search of gaugings could be easily performed. As $SU(8)$ is trivial, we focus on $SL'(8)$ and as in the previous section we will parametrize the frame in terms of matrices in the fundamental representation. To do that we introduce the following $Sp(56,{\mathbb R})$ rotation matrix,
\bea
R=\frac{\sqrt{2}}{4}\left(\begin{matrix}\Gamma_{\breve{A}\breve{B}}{}^{AB}&\Gamma_{\breve{A}\breve{B}~ AB} \cr
                                       -\Gamma_{\breve{A}\breve{B}~AB}&\Gamma^{\breve{A}\breve{B}}{}_{AB}\end{matrix}\right)~,\label{RotSL'(8)}
\eea
wherein the chiral gamma matrices, which intertwine between both $SL(8)$ indices, are chosen to satisfy the duality relations (\ref{dualityG}). A rotation of the frame with (\ref{RotSL'(8)}) exchanges the role of $SL'(8)$ and $SL(8)$ subgroups, in particular the former arises as the block diagonal realized. Here we do not want to change the frame but we can use this rotation to parametrize a general bein from $SL'(8)$ as
\bea
U'{}^{\mathbb A}{}_{\mathbb M}=\Gamma^{\breve{A}\breve{B}}_{AB}\Gamma^{\breve{M}\breve{N}}_{MN}
\left(\begin{matrix} U{}_{[\breve{A}}{}^{\breve{M}}U{}_{\breve{B}]}{}^{\breve{N}} + U{}^{[\breve{A}}{}_{\breve{M}}U^{\breve{B}]}{}_{\breve{N}} &
U{}_{[\breve{A}}{}^{\breve{M}}U{}_{\breve{B}]}{}^{\breve{N}} - U{}^{[\breve{A}}{}_{\breve{M}}U^{\breve{B}]}{}_{\breve{N}}                      \cr
U{}_{[\breve{A}}{}^{\breve{M}}U{}_{\breve{B}]}{}^{\breve{N}} - U{}^{[\breve{A}}{}_{\breve{M}}U^{\breve{B}]}{}_{\breve{N}}                      &
U{}_{[\breve{A}}{}^{\breve{M}}U{}_{\breve{B}]}{}^{\breve{N}} + U{}^{[\breve{A}}{}_{\breve{M}}U^{\breve{B}]}{}_{\breve{N}}\end{matrix}\right)~,\label{U'}
\eea
where $U{}^{\breve{A}}{}_{\breve{M}}$ is in the ${\bf 8}$ representation and $U{}_{\breve{A}}{}^{\breve{M}}$ denotes its inverse transposed, so they still satisfy (\ref{unimodular}). The vanishing of the {\bf 56} gaugings $\vartheta_{\mathbb A}$ now is given by
\bea
\Gamma^{\breve{M}\breve{N}}_{MN}~ U_{[\breve{A}}{}^{\breve{M}}(\partial_{MN}-\partial^{MN})\left( U_{\breve{B}]}{}^{\breve{N}}e^{-\frac{3\Delta}2}\right)=0~,~~~~ 
\Gamma^{\breve{M}\breve{N}}_{MN}~ U^{[\breve{A}}{}_{\breve{M}}(\partial_{MN}+\partial^{MN}) \left(U^{\breve{B}]}{}_{\breve{N}}e^{-\frac{3\Delta}2}\right)=0~.~~\label{no56'rep}
\eea
Inserting (\ref{U'}) in (\ref{Xtensor}), and using the equations (\ref{unimodular}) and (\ref{no56'rep}) we obtain the projections on ${\bf 36}$ and ${\bf 420}$,
\bea
\theta_{AB}&=&
-~\delta_{AB}~e^{-\Delta}~ \Gamma^{MN}_{{\breve{M}\breve{N}}} \left[U_{\breve{A}}{}^{\breve{M}} \partial_{MN} U_{\breve{A}}{}^{\breve{N}} + U^{\breve{A}}{}_{\breve{M}} \partial_{MN}U^{\breve{A}}{}_{\breve{N}}  \right.\cr
&&\left. ~~~~~~~~~~~~~~~~~~~~~~~~~~~~~~~~~~~~~~+ 
U^{\breve{A}}{}_{\breve{M}} \partial^{MN}U^{\breve{A}}{}_{\breve{N}}-
U_{\breve{A}}{}^{\breve{M}} \partial^{MN} U_{\breve{A}}{}^{\breve{N}} \right] \cr
&&-
\frac12~\Gamma^{\breve{A}\breve{B}\breve{C}\breve{D}}_{AB}~e^{-\Delta}~ \Gamma^{MN}_{{\breve{M}\breve{N}}} 
\left[f^{+~\breve{A}\breve{B}}_{\breve{M}\breve{N}} 
U^{\breve{C}}{}_{\breve{Q}}\partial_{MN} U_{\breve{D}}{}^{\breve{Q}} 
+ f^{-~\breve{A}\breve{B}}_{\breve{M}\breve{N}} U^{\breve{C}}{}_{\breve{Q}}\partial^{MN} U_{\breve{D}}{}^{\breve{Q}}\right]~,
\cr\cr
\zeta_{A}{}^{BCD}&=&
       \frac18~ e^{-\Delta}~
\left(18~ \delta_{A}^{[B}\Gamma^{CD]}_{\breve{A}(\breve{C}}\delta_{\breve{D})}^{\breve{B}}- \Gamma^{EA}_{\breve{A}\breve{B}}~ \Gamma^{BCDE}_{\breve{C}\breve{D}}\right)~ \Gamma^{MN}_{\breve{M}\breve{N}}\cr
&&~~~~~~~~~~~~~~~~~~~~~~~~~~~~~~~\times~ \left(f^{+~\breve{A}\breve{B}}_{\breve{M}\breve{N}}
U^{\breve{C}}{}_{\breve{Q}}\partial_{MN} U_{\breve{D}}{}^{\breve{Q}} + 
f^{-~\breve{A}\breve{B}}_{\breve{M}\breve{N}}
U^{\breve{C}}{}_{\breve{Q}}\partial^{MN} U_{\breve{D}}{}^{\breve{Q}}\right)\cr
       &+&~\frac3{16}~ e^{-\Delta}~ \left(\Gamma^{[BC}_{\breve{A}\breve{B}}\Gamma^{D]A}_{\breve{C}\breve{D}}-8\delta^{\breve{A}}_{\breve{[C}}\Gamma_{\breve{D}]\breve{B}}^{[BC}\delta^{D]}_{A}\right)~ \Gamma^{MN}_{\breve{M}\breve{N}}\cr
&&~~~~~~~~~~~~~~~~~~~~~~~~~~~~~~~\times~ \left(f^{-~\breve{A}\breve{B}}_{\breve{M}\breve{N}}
U^{\breve{C}}{}_{\breve{Q}}\partial_{MN} U_{\breve{D}}{}^{\breve{Q}} + 
f^{+~\breve{A}\breve{B}}_{\breve{M}\breve{N}}
U^{\breve{C}}{}_{\breve{Q}}\partial^{MN} U_{\breve{D}}{}^{\breve{Q}}\right)\cr
       &+&~\frac{15}8~ e^{-\Delta}~
\delta_{A}^{[B}\Gamma^{CD]}_{\breve{A}\breve{B}}~ \Gamma^{MN}_{\breve{M}\breve{N}}
\left(f^{-~\breve{A}\breve{B}}_{\breve{M}\breve{N}}
\partial_{MN}\Delta + 
f^{+~\breve{A}\breve{B}}_{\breve{M}\breve{N}}
\partial^{MN}\Delta\right)~,\label{SL'8gaugings} 
\eea
where $f^{\pm~\breve{A}\breve{B}}_{\breve{M}\breve{N}}=\left(U^{\breve{A}}{}_{\breve{M}} U^{\breve{B}}{}_{\breve{N}} \pm U_{\breve{A}}{}^{\breve{M}}U_{\breve{B}}{}^{\breve{N}} \right)$. With the help of (\ref{GammaId}) one sees that the latter satisfies $\zeta_A{}^{ACD}=0$, as it should. The projections on ${\bf 36'}$ and ${\bf 420'}$ coincide with the unprimed ones after $\partial_{MN}\leftrightarrow\partial^{MN}$ up to a global sign.

We are now in a position to explore inside this truncation, for instance we can ask if this sector admits dyonic $SO(8)$ solutions or not. 
With the lack of a consistent way to relax the {\it section conditions} let us consider a situation where they hold and without loss of generality let us assume the physical slice is given by $y^m=X^{m8}$. If such a solution exists, we can always consider a basis where $\theta_{AB}=\lambda_+\delta_{AB},~\xi^{AB}=\lambda_-\delta^{AB}$, so that we end with the following system of equations
\bea
e^{-\Delta}~\Gamma^m_{\breve M\breve N}~U_{\breve A}{}^{\breve M}\partial_m U_{\breve A}{}^{\breve N}&=&K_1~,~~~~~~~~~~~
e^{-\Delta}~\Gamma^m_{\breve M\breve N}~U^{\breve A}{}_{\breve M}\partial_m U^{\breve A}{}_{\breve N}=K_2~,~~~~\cr
\Gamma^m_{\breve M\breve N}~\Gamma^{\breve A\breve B\breve C \breve D}_{AB}~U_{\breve A}{}^{\breve M} U_{\breve B}{}^{\breve N} U_{\breve C}{}^{\breve Q}\partial_m U^{\breve D}{}_{\breve Q}&=&0~,~~~~
\Gamma^m_{\breve M\breve N}~\Gamma^{\breve A\breve B\breve C \breve D}_{AB}~U^{\breve A}{}_{\breve M} U^{\breve B}{}_{\breve N} U^{\breve C}{}_{\breve Q}\partial_m U_{\breve D}{}^{\breve Q}=0~.~~~~~~~~~\label{36gaugings}
\eea
In order to have non vanishing dyonic gaugings, $\lambda_\pm=-\frac12\left(K_1\pm K_2\right)\neq0$. Contracting last two lines with $\Gamma^{\breve A'\breve B'\breve C'\breve D'}_{AB}$ we see that only their self-dual parts are constrained
\bea
P^{(+)}{}^{\breve A~\breve B~ \breve C ~\breve D}_{\breve A' \breve B' \breve C' \breve D'} ~\Gamma^m_{\breve M\breve N}~ U_{\breve A}{}^{\breve M} U_{\breve B}{}^{\breve N} U_{\breve C}{}^{\breve Q}\partial_m U^{\breve D}{}_{\breve Q}&=&0~,\cr
P^{(+)}{}^{\breve A~\breve B~ \breve C~ \breve D}_{\breve A' \breve B' \breve C' \breve D'}~\Gamma^m_{\breve M\breve N}~ U^{\breve A}{}_{\breve M} U^{\breve B}{}_{\breve N} U^{\breve C}{}_{\breve Q}\partial_m U_{\breve D}{}^{\breve Q}&=&0~,~\label{P+projection}
\eea
where we have introduced the (anti) self-duality projectors 
\bea
P^{(\pm)}{}^{\breve A~\breve B~ \breve C~ \breve D}_{\breve A' \breve B' \breve C' \breve D'}=\frac12\left(4!~ \delta^{\breve A}_{[\breve A'}\delta^{\breve B}_{\breve B'}\delta^{\breve C}_{\breve C'} \delta^{\breve D}_{\breve D']} \pm \epsilon_{\breve A'\breve B'\breve C'\breve D'\breve A\breve B\breve C\breve D}\right)~.
\eea
We also have to impose that ${\bf 420}$ and ${\bf 420'}$ are projected out, which leads to
\bea
&& \left(6 \delta_{A}^{[B}\Gamma^{CD]}_{\breve{A}(\breve{C}}\delta_{\breve{D})}^{\breve{B}}- \Gamma^{EA}_{\breve{A}\breve{B}}~ \Gamma^{BCDE}_{\breve{C}\breve{D}}+\frac3{2}~ \Gamma^{[BC}_{\breve{A}\breve{B}}\Gamma^{D]A}_{\breve{C}\breve{D}}\right)~ \Gamma^{m}_{\breve{M}\breve{N}}
 U_{\breve{A}}{}^{\breve{M}}U_{\breve{B}}{}^{\breve{N}}
U_{\breve{C}}{}^{\breve{Q}}\partial_{m} U^{\breve{D}}{}_{\breve{Q}} \cr
&&~~~~~~~~~~-2~ \delta_{A}^{[B}\Gamma^{CD]}_{\breve{A}\breve{B}}~ \Gamma^{m}_{\breve{M}\breve{N}}
U_{\breve{A}}{}^{\breve{M}}\partial_m~ U_{\breve{B}}{}^{\breve{N}}=0~,\cr
&& \left(6 \delta_{A}^{[B}\Gamma^{CD]}_{\breve{A}(\breve{C}}\delta_{\breve{D})}^{\breve{B}}- \Gamma^{EA}_{\breve{A}\breve{B}}~ \Gamma^{BCDE}_{\breve{C}\breve{D}}+\frac3{2}~ \Gamma^{[BC}_{\breve{A}\breve{B}}\Gamma^{D]A}_{\breve{C}\breve{D}}\right)~ \Gamma^{m}_{\breve{M}\breve{N}}
 U^{\breve{A}}{}_{\breve{M}}U^{\breve{B}}{}_{\breve{N}}
U^{\breve{C}}{}_{\breve{Q}}\partial_{m} U_{\breve{D}}{}^{\breve{Q}} \cr
&&~~~~~~~~~~-2~ \delta_{A}^{[B}\Gamma^{CD]}_{\breve{A}\breve{B}}~ \Gamma^{m}_{\breve{M}\breve{N}}
U^{\breve{A}}{}_{\breve{M}}\partial_m~ U^{\breve{B}}{}_{\breve{N}}=0~. \label{420gaugings}
\eea

These are a priori ${\bf 420}+{\bf 420'}$ equations and the compatibility with (\ref{36gaugings}) is far from being evident. To read more clearly some of the constraints imposed by (\ref{420gaugings}) we can make some appropriate contractions with gamma matrices. For instance, by multiplying (\ref{420gaugings}) with $\Gamma^{[BC}_{[\breve A' \breve B'}\Gamma^{D]A}_{\breve C' \breve D']}$ we get 
\bea
P^{(-)}{}^{\breve A~\breve B~ \breve C ~\breve D}_{\breve A' \breve B' \breve C' \breve D'} ~\Gamma^m_{\breve M\breve N}~ U_{\breve A}{}^{\breve M} U_{\breve B}{}^{\breve N} U_{\breve C}{}^{\breve Q}\partial_m U^{\breve D}{}_{\breve Q}&=&0~,\cr
P^{(-)}{}^{\breve A~\breve B~ \breve C~ \breve D}_{\breve A' \breve B' \breve C' \breve D'}~\Gamma^m_{\breve M\breve N}~ U^{\breve A}{}_{\breve M} U^{\breve B}{}_{\breve N} U^{\breve C}{}_{\breve Q}\partial_m U_{\breve D}{}^{\breve Q}&=&0~,
\eea
where we have repeatedly used relations (\ref{GammaId}). Plugging it into (\ref{P+projection}) leads to 
\bea
\Gamma^m_{[\breve M\breve N}~U^{\breve A}{}_{\breve P}\partial_{|m|}U^{\breve A}{}_{\breve Q]}=0~~,~~~~~~
\Gamma_m^{[\breve M\breve N}~U_{\breve A}{}^{\breve P}\partial_{m}U_{\breve A}{}^{\breve Q]}=0~.\label{P+P-Constraint}
\eea
Of course there is much more information in (\ref{420gaugings}), but we do not need it because we are now in a position to see that (\ref{P+P-Constraint}) is incompatible with the first two equations in (\ref{36gaugings}). In fact, contraction of (\ref{P+P-Constraint}) with $\Gamma^n_{MN}\Gamma^n_{PQ}$ yields $K_1=K_2=0$.

\section{Discussions and Conclusions}\label{secConcl}

In this work we have addressed the challenge of embedding tensor building from the U-dual Extended Geometry perspective by concrete realizations of generalized Scherk-Schwarz reductions. This approach offers an attractive mechanism in the search of gaugings with a higher dimensional interpretation. In particular each embedding tensor leads to a differential system of equations for the so called generalized vielbeins, $i.e.$ $E_7$ valued objects encoding the internal data, $e.g.$ geometric and non-geometric fluxes. Unfortunately, $E_7$ is very involved to carry out this strategy in a straightforward way and a systematic exploration is requested. Following this program we performed the first step by analyzing the situation where the generalized bein is in one of the maximal nontrivial subgroups. These correspond to the two inequivalent SL(8) subgroups and the embedding tensor projections on the irreducible representations {\bf 36,~36',~420} and {\bf 420'} of the $sl(8)$ branching are displayed in (\ref{SL8gaugings}) and (\ref{SL'8gaugings}) respectively. 

Then we probed the handy of these expressions by focusing on the $SO(8)$ gaugings. For instance we proved that neither of these sectors yield dyonic gaugings and we showed how the generalized frame leading to the electric ones can be recovered without any reference to the fermionic sector as opposed to the ground up approach of Nicolai {\it et al.} 

The analysis presented here was performed purely from the algebraic point of view, meaning that the only ingredient used so far was the generalized diffeomorphism (\ref{GenDiff}) and so applying these results to a concrete theory like the Exceptional Field Theory of Hohm and Samtleben \cite{Hohm:2013pua,Hohm:2013uia} has the additional requirement that the 4+56 dependent frame must be a solution for the action. 

There are several possible extensions of this work. For instance, the obvious next step is to extend the analysis to more general $E_7$ beins, not restricted to the maximal subgroups. In particular it could be used in the program for understanding if there is an uplift to 11D supergravity/M theory for the new $SO(8)_c$ gauged supergravities described in \cite{DallAgata:2011aa,Dall'Agata:2012bb}. 

Another aspect deserving further attention is the possibility of relaxing the {\it section conditions}, even though in this paper we assumed their validity in the search of concrete $SO(8)$ solutions, neither the truncation on $SL(8)$ (equation (\ref{SL8gaugings})) nor the one on $SL'(8)$ (equation (\ref{SL'8gaugings})) was performed by explicit use of the section constraints. Therefore, the expressions obtained in this work are expected to be relevant also for a speculative situation where they are weakened.
  
Finally, another interesting subject is the construction of $SO(p,8-p)$ or more generally $CSO(p,q,8-p-q)$ gaugings withing this formalism. There are known concrete realizations of some of these gauged supergravities (see for instance \cite{Hull:1988jw}) from 11D. We expect that the Extended Geometry approach can shed light on the construction of generalized beins for the non-compact groups, leading to an $E_{7}$ covariant reformulation of 11D supergravity as was done for the seven sphere as well as for twisted tori backgrounds \cite{Godazgar:2013oba}. In addition, this can also be used to explore about possible uplifts of the non-compact dyonic gauged supergravities. These situations are currently under consideration and constitute part of a subsequent paper. 

\bigskip

{\bf \large Acknowledgments} {We are very grateful to Diego Marques, Gianguido Dall'Agata and Erik Plauschinn for useful discussions and valuable comments on the manuscript. We also thank to Hermann Nicolai, Mahdi Godazgar for correspondence. This work was supported by MIUR grant RBFR10QS5J.}

\begin{appendix}
\section{Appendix}
{\Large \bf {$E_7$ in the SL(8) frame}}
\medskip

Along this work we use the $sl(8)$ branching of the exceptional $e_{7(7)}$ algebra, which split the adjoint and fundamental representations as
\newpage
\bea
{\bf 133} &\longrightarrow& {\bf 63} + {\bf 70}~,\cr
{\bf 56} &\longrightarrow& {\bf 28} + {\bf 28'}~.
\eea

In this branching the elements of the algebra are parametrized by 
\bea
\left(\begin{matrix}\Lambda^{ij}{}_{kl}& \Sigma^{ijkl}\cr\Sigma_{ijkl}&\Lambda_{ij}{}^{kl}\end{matrix}\right),
\eea
with arbitrary real antisymmetric tensors $\Sigma_{ijkl}$ and $\Sigma^{ijkl}=\frac1{24}\epsilon^{ijklmnpq}\Sigma_{mnpq}$, where $\epsilon_{12345678}=\epsilon^{12345678}=1$. $\Lambda^{ij}{}_{kl}$ defined in term of the traceless matrices $\lambda^i{}_k$ as in (\ref{SL(8)}) parametrizes $sl(8)$, one of the three maximal subalgebras of $e_7$. The other two are $su(8)$ and another $sl(8)$. All of them shear the same $so(8)$ subalgebra generated by the antisymmetric $\lambda^i{}_j$. The unitary algebra is generated by complementing them with anti self-duals $\Sigma_{ijkl}$ and the other $sl(8)$ by complementing them with the self-dual ones.

The generators $t_\alpha$ split in the 63 of $sl(8)$, $t^A{}_B$ and the 70 $t^{ABCD}$, whose non-vanishing components are
\bea
\left[t^{A}{}_{B}\right]^{[CD]}{}_{[EF]}&=&-\left[t^{A}{}_{B}\right]_{[EF]}{}^{[CD]}
=-\left(\delta^{A}_{[E}\delta^{[C}_{F]}\delta^{D]}_{B}+\frac18 \delta^{A}_{B}\delta^{C}_{[E}\delta^{D}_{F]}\right)~,
\eea
\bea
\left[t^{A B C D}\right]^{[EF]\; [GH]}&=&\frac1{24}\epsilon^{A B C D E F G H}~,~~
\left[t^{A B C D}\right]_{[EF]\; [GH]}=\delta^{A}_{[E}\delta^{B}_{F}\delta^{C}_{G}\delta^{D}_{H]}~,
\eea
and the Killing metric, $K_{\alpha\beta}$ is block diagonal
\bea
K^{A_1}{}_{A_2},{}^{B_1}{}_{B_2}&=&\frac34\left(\delta^{A_1}_{B_2}\delta^{B_1}_{A_2}-\frac18\delta^{A_1}_{A_2}\delta^{B_1}_{B_2}\right)~,\cr
K_{A_1 A_2 A_3 A_4\,,\,B_1 B_2 B_3 B_4}&=&48~ \epsilon_{A_1 A_2 A_3 A_4 B_1 B_2 B_3 B_4}~.
\eea

The definition of the generalized diffeomorphisms and the embedding tensor use the projectors to the adjoint and ${\bf 912}$ representations
\bea
P_{(adj)}{}^{A}{}_{B}{}^{C}{}_{D}=\left[ t_\alpha\right]_B{}^A~\left[ t^\alpha\right]_D{}^C~,\label{padj}
\eea
\bea
P_{(912)A}{}^{\alpha},{}^{B}{}_{\beta}=\frac17\delta^\alpha_\beta \delta_A^B-\frac{12}{7}\left[ t_\beta t^\alpha\right]_A{}^B
+\frac{4}{7}\left[t^\alpha t_\beta\right]_A{}^B~,\label{p912}
\eea
where $t^\alpha$ are obtained by raising $\alpha$ with the inverse Killing metric. Of course they satisfy $P_{(adj)}{}^{\mathbb A}{}_{\mathbb B}{}^{\mathbb B}{}_{\mathbb A}=133$ and $P_{(912)}{}_{\mathbb A}{}^{\alpha,\mathbb A}{}_{\alpha}=912$~.
\bigskip
\bigskip

{\Large \bf Gamma matrices}
\medskip

Our conventions for the gamma matrices are the following. The Clifford algebra is 
\bea
\left\{\Gamma_a,\Gamma_b\right\}_{\breve{A}}{}^{\breve{B}}=2~\delta_{ab} ~\delta_{\breve{A}}^{\breve{B}}~,\label{Clifford}
\eea
$a,b=1,\dots,7;~\breve A,\breve B=1,\dots,8$. Both kind of indices are raised and lowered with the Kronecker's delta, so their position is meaningless. The completely antisymmetric gamma matrices with indices $A=\{a,8\}$ is defined as $\Gamma_{ab}=\Gamma_{[a}\Gamma_{b]}$, and $\Gamma_{a8}=i\Gamma_{a},~a,b=1,\dots,7$. 

Some useful relations we used in this work are 
\bea
\left\{\Gamma_{AB},\Gamma_{CD}\right\}&=&-4~\delta_{[A}^{C}\delta_{B]}^{D}~{\mathbb I} + 2 ~\Gamma_{ABCD}~,\cr
\left[\Gamma_{AB},\Gamma_{CD}\right]&=&4~\left(\delta_{A[C}\Gamma_{D]C}-\delta_{B[C}\Gamma_{D]A}\right)~.\label{GammaId}
\eea
In addition, we choose the matrices with the following selfduality relations 
\bea
[\Gamma_{ABCD}]_{\breve{A}}{}^{\breve{B}}&=&\frac{1}{4!}~\epsilon_{ABCDEFGH}[\Gamma^{EFGH}]_{\breve{A}}{}^{\breve{B}}.\label{dualityG}
\eea
Finally we introduce the $\Gamma_{\breve A\breve B}$ matrices via
\bea
\left[\Gamma_{\breve{A}\breve{B}}\right]_{A}{}^{B}:=\left[\Gamma_{AB}\right]_{\breve{A}}{}^{\breve{B}}~.
\eea
These satisfy the same relations (\ref{GammaId})-(\ref{dualityG}) by exchanging hated and unhated indices. Because of this symmetry we denote them along the work simply as $\Gamma_{\breve{A}\breve{B}}^{AB}$ or by any variation in the position of their indices, which as we said is meaningless.

\end{appendix}

\end{document}